\documentclass[iop]{emulateapj}

\usepackage{multirow} 
\usepackage{hyperref}

\def\lapp{\ifmmode\stackrel{<}{_{\sim}}\else$\stackrel{<}{_{\sim}}$\fi}
\def\gapp{\ifmmode\stackrel{>}{_{\sim}}\else$\stackrel{>}{_{\sim}}$\fi}

\newcommand{\source}{1RXS~J170849.0$-$400910}
\newcommand{\src}{RXS~J1708}
\newcommand{\rxte}{\textit{RXTE}}
\newcommand{\xte}{\textit{RXTE}}
\newcommand{\cxo}{\textit{Chandra}}
\newcommand{\chandra}{\textit{Chandra}}
\newcommand{\xmm}{\textit{XMM}}
\newcommand{\swift}{\textit{Swift}}
\newcommand{\rosat}{\textit{ROSAT}}
\newcommand{\asca}{\textit{ASCA}}
\newcommand{\sax}{\textit{BeppoSAX}}
\newcommand{\integral}{\textit{INTEGRAL}}

\newcommand{\degrees}{^{\circ}}
\newcommand{\kes}{PSR~J1846$-$0258}
\newcommand{\meighteen}{1E~1841$-$045}
\newcommand{\mten}{1E~1048.1$-$5937}
\newcommand{\mohone}{4U~0142+61}
\newcommand{\mtwenty}{1E~2259+586}

\begin{document}

\title{On the X-ray variability of magnetar \source }

\author{P. Scholz\altaffilmark{1,6}, 
R. F. Archibald\altaffilmark{1},
V. M. Kaspi\altaffilmark{1}, 
C.-Y. Ng\altaffilmark{1,2} 
A. P. Beardmore\altaffilmark{3}, 
N. Gehrels\altaffilmark{4}, 
and J. A. Kennea\altaffilmark{5}}
%

\altaffiltext{1}{Department of Physics, Rutherford Physics Building, McGill University, 3600 University Street, Montreal, Quebec, H3A 2T8, Canada}
\altaffiltext{2}{Department of Physics, The University of Hong Kong, Pokfulam Road, Hong Kong}
\altaffiltext{3}{Department of Physics and Astronomy, University of Leicester, University Road, Leicester LE1 7RH, UK}
\altaffiltext{4}{Astrophysics Science Division, NASA Goddard Space Flight Center, Greenbelt, MD 20771 USA}
\altaffiltext{5}{Department of Astronomy and Astrophysics, 525 Davey Lab, Pennsylvania State University, University Park, PA 16802, USA}
\altaffiltext{6}{Contact email: pscholz@phsyics.mcgill.ca}

\begin{abstract}

We present a long-term X-ray flux and spectral analysis for \source\ using \swift/XRT spanning over 8 years
from 2005--2013. We also analyze two observations from \chandra\ and \xmm\ in the period from 2003--2004.
In this 10-yr period, \source\ displayed several rotational glitches. Previous studies have claimed
variations in the X-ray emission associated with some of the glitches. From our analysis we find
no evidence for significant X-ray flux variations and evidence for only low-level spectral variations. 
We also present an updated
timing solution for \source, from \xte\ and \swift\ observations, 
which includes a previously unreported glitch at MJD 56019.
We discuss the frequency and implications of radiatively quiet glitches in magnetars.
\end{abstract}


\section{Introduction}

Magnetars are a type of pulsar that exhibit exotic and often violent properties. Their defining characteristic
is that they are powered not by their rotation, as are most Crab-like pulsars, but by the decay of their high
magnetic fields. Because of the energy provided by the magnetic field decay, their X-ray luminosities are 
generally higher than their rotational spin-down energies.
They often display outburst activity during which they can increase their brightness
by an order of magnitude or more and emit short ($\sim10$\,ms to $\sim1$\,s in duration) energetic bursts. Previously,
magnetars had been classified into two observational categories: Anomolous X-ray pulsars (AXPs) and 
Soft-gamma repeaters (SGRs). However, these two ``classes" appear to be merely different ends of the magnetar
behavioral spectrum. 
For a review see \citet{wt06}.

Glitches in magnetars have been observed both with and without associated radiative changes. Out of the 26 magnetars
and magnetar candidates\footnote{See the magnetar catalog at \url{http://www.physics.mcgill.ca/$\sim$pulsar/magnetar/main.html}}
\citep{ok13}
 only five are monitored sufficiently frequently to detect unambiguously the occurance of glitches \citep{dk13}. 
Of those five, one, \meighteen, has never displayed any radiative activity associated with its glitches \citep{zk10} whereas
\mten, \mtwenty, and \mohone\ have had radiative events during some or all of their glitches \citep{dkg09,kgw+03,gdk11,dk13}.
It is important to determine whether or not there is a generic connection between magnetar glitches and
radiative events because it can help us determine the physical orgin of these phenomena. 
It seems reasonable that magnetospheric mechanisms, 
because of their
external nature, are likely to be accompanied by radiative changes whereas internal mechanisms could produce
radiatively quiet glitches.

\source\ (hereafter referred to as \src\ for brevity) was first identified as an X-ray source in the \rosat\ all-sky survey \citep{vab+99}. 
It was first discovered as a pulsar by \citet{snt+97}, using \asca\ data, 
who suggested that it was an AXP based on its X-ray spectrum and 11-s spin period. 
\citet{ics+99} measured a period derivative typical of AXPs for \src, 
confirming that the source is an AXP, and thus a magnetar.

\src\ was the first magnetar observed to glitch \citep{klc00}.
It has since been found to glitch several more times \citep{igz+07,dkg08,dk13}. 
Note that some of the glitches reported in \citet{igz+07} are considered
to be glitch candidates in \citet{dkg08} as they could be consistent with timing noise.
\citet{roz+05} first suggested
that \src\ exhibited post-glitch X-ray flux variability based on a 2003 \xmm\ observation. 
They reported an \xmm\ flux that was significantly lower than preceding \cxo\ 
and \sax\ observations. Futher evidence for flux variability was claimed based on
 additional \swift\ and \integral\ observations \citep{cri+07,gri+07,igz+07}.
However, puzzlingly, variability at the level claimed in these studies was not seen the pulsed
count rate as measured by frequent observations with \rxte\ \citep{dkg08,dk13}.

In this paper we analyze all the available \swift\ X-Ray Telescope (XRT) 
data from the period of the claimed variability to present day. We also use
 one \xmm\ and one \chandra\ observation that were performed prior to the start
of the \swift\ observations. We then use the measured spectral and flux values to constrain
the level of source variability. We also present an up-to-date timing solution which 
continues the \xte\ timing of \citet{dk13} using \swift. We then discuss the occurance 
of radiatively quiet glitches in magnetars.


\section{Observations}

\subsection{\swift\ Observations}

\src\ was observed by the \swift\ XRT frequently between 
2005 and 2010. Beginning in July 2011 \src\ was observed as a continuation
of the \xte\ timing campaign summarized by \citet{dk13}. 
Here we use all available archival \swift\ data in that time period
in both Windowed Timing (WT) and Photon Counting (PC) modes. 
There were 80 observations for a total exposure time of 268\,ks. Table \ref{ta:obs}
shows a summary of the \swift\ observations used in this work.

We downloaded the unfiltered Level 1 data from the {\em HEASARC} data archive
and ran the standard \swift\ data reduction script {\ttfamily xrtpipeline} using
the source position of 
$17^{\rm{h}}$~$08^{\rm{m}}$\,$46.87^{\rm{s}}$, 
$-40\degrees$\,$08\arcmin$\,$52.44\arcsec$ \citep{icp+03}
and the best available spacecraft attitude file.
Events were then reduced to the solar-system barycenter using the same position.
For WT mode, a 30-pixel long strip centered on the source was used to extract the 
source events and a 50-pixel long strip positioned away from the source was
used to extract the background events.
For PC mode observations, an annular region with inner radius 3 pixels and outer radius
20 pixels was used. The inner region was excluded to avoid
pileup of the source. An annulus with inner radius 40 pixels and outer 
radius 60 pixels was used as the background region.

For WT mode data, exposure maps, spectra, and ancillary response files
were created for each individual orbit. The spectra and ancillary response
files were then summed to create a spectrum for each observation. For
the PC mode data, exposure maps, spectra and ancillary response files
were created on a per observation basis. We used response files for spectral fitting
from the 20120209 CALDB. 

The use of exposure maps when creating the ancillary response files is especially
important for \swift\ data, as there are columns of bad pixels which can
disrupt the PSF of the source for parts of certain observations. Orbits
were not used in the observation if the bad columns were found to be 
within 3 pixels of the source position. 

For WT mode data we selected only Grade 0 events for spectral fitting as
higher Grade events are more likely to be caused by a background event \citep{bhn+05}. 
In PC mode we used the standard Grade 0-12 selection.


\subsection{\cxo\ and \xmm\ Observations} 
\label{sec:cxoobs}

In this study, we also reprocessed archival data taken with the {\em Chandra
X-ray Observatory} and {\em XMM-Newton}. To avoid pileup, we used the \cxo\
continuous-clocking (CC) mode observation (ObsID 4605) and the \xmm\ PN
small-window mode data (ObsID 0148690101). The former was taken on 2004 July
3 with the ACIS-S detector in CC mode, which has a time resolution of 3\,ms.
The total exposure was 29\,ks. The \xmm\ observations were made on 2003
August 28. The PN and MOS detectors were run in small and large window modes,
with 0.5-s and 6-ms time resolution, respectively. As the source is bright,
the low time resolution of the MOS data results in significant pileup.
Therefore, we focused only on the PN data. After filtering for periods of high
background, we were left with 35\,ks of exposure. This is equivalent to 24\,ks of live
time since the small-window mode has an efficiency of 70\%.

We processed the \cxo\ and \xmm\ data using CIAO 4.4 and SAS 11,
respectively. The source spectrum was extracted using a 6\arcsec-wide region
from the \cxo\ observation and a 40\arcsec-radius aperture from the
\xmm\ PN data. For the \cxo\ observation, 
the background spectrum was extracted from the entire 1D CC-mode
strip excluding the inner 1\arcmin\ closest to the source. For the \xmm\ observation,
the background spectrum was extracted from two 40\arcsec-radius circular regions
placed away from the source.

\section{Analysis \& Results}

\subsection{Flux and spectra}
\label{sec:spec_analysis}

\begin{figure}
\plotone{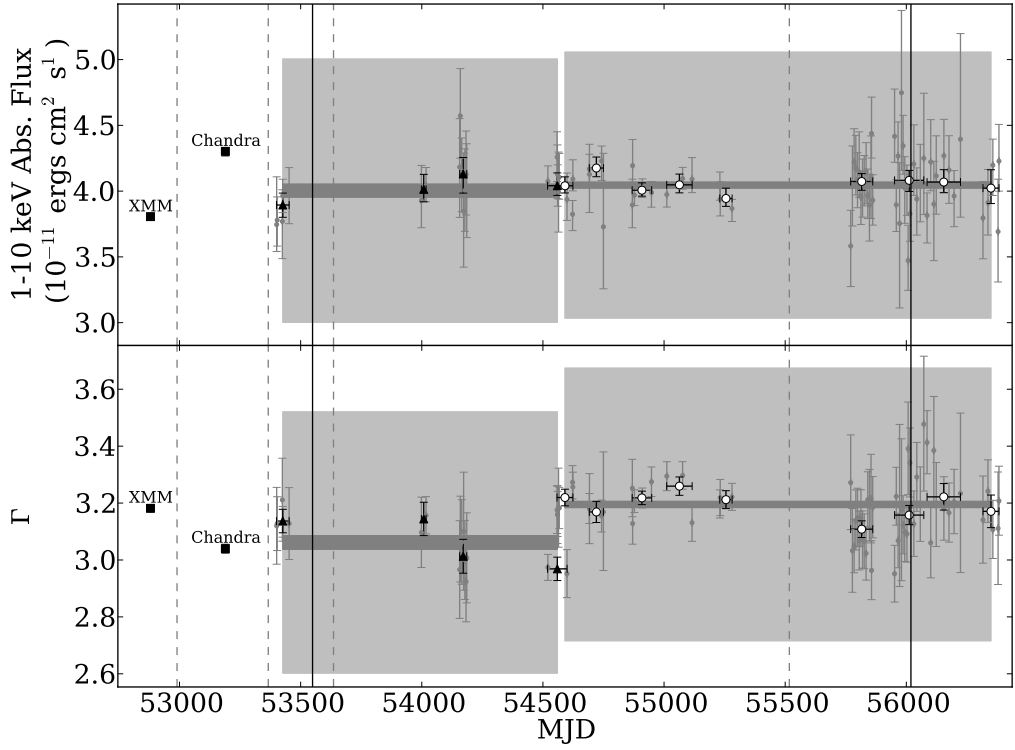}
\figcaption{
{\em Top Panel:} Absorbed 1--10\,keV flux of \src\ over a $\sim$10-yr period. Note that
the zero on the y-axis is suppressed. 
{\em Bottom Panel:} Photon indices from fitting a power-law model to the 1--10\,keV spectrum.
Grey points are from spectral fits to individual observations.
Black triangles are sets of \swift\ PC mode observations, and white circles are sets
of \swift\ WT mode observations (see Table \ref{ta:obs} for definitions of sets).
\xmm\ and \cxo\ observations are labelled.
The dark grey bands represent the 90\% error in the mean and the light grey bands
represent the level of previously claimed variabilty ($\sim50$\% in 1--10\,keV flux and
$\sim30$\% in spectral index; \citealt{gri+07}). The solid vertical lines represent the epochs of glitches
and the dashed lines indicate the epochs of glitch candidates. All error bars are 90\% confidence intervals.
\label{fig:flux}
}
\end{figure}

We first fit the spectra for each individual observation with a photoelectrically absorbed power-law model.
The spectra were fit
with a single $N_{\mathrm H}$ using the XSPEC {\ttfamily tbabs} model with abundances from \citet{wam00}, and 
photoelectric cross-sections from \citet{vfky96}.
We used XSPEC\footnote\url{http://xspec.gfsc.nasa.gov} with Cash statistics \citep{cas79a} to fit the spectra
because of the low number of counts in the \swift\ observations.
The grey points in Figure \ref{fig:flux} show the results of the spectral fits to the individual observations.
The typical uncertainties in the spectral parameters vary widely due to the large range in exposure times.
In order to place the best constraints on the variability, we consider PC and WT modes separately.
This is because the two modes are calibrated to within only 10\% of each other (A. Beardmore, private communication).
The mean and standard deviation of the 1--10\,keV absorbed flux for the PC mode data are $4.0\times10^{-11}$\,erg\,cm$^{-2}$\,s$^{-1}$ and
$1.9\times10^{-12}$\,erg\,cm$^{-2}$\,s$^{-1}$, respectively. The PC mode photon index has a mean and standard deviation of
3.1 and 0.08. For WT mode the mean and standard deviation are $4.0\times10^{-11}$\,erg\,cm$^{-2}$\,s$^{-1}$ and
$2.1\times10^{-12}$\,erg\,cm$^{-2}$\,s$^{-1}$ for flux and 3.2 and 0.07 for the photon index.

In order to better constrain the variability,
we then separated the \swift\ spectra into sets of observations nearby in time (see Table \ref{ta:obs}).
Within each set, the 1--10\,keV flux and photon-index were consistent with being constant (i.e. the $\chi^2$ values of fits
to a mean value in each set were consistent with being drawn from a $\chi^2$ distribution). 
We fit the sets of \swift\ observations as well as the  \xmm\ and \cxo\ spectra with a photoelectrically absorbed power law. 
Each \swift\ set was fitted with the same model with all spectral parameters the same from observation to observation within 
the set.
All parameters were allowed to vary from set to set except for $N_{\mathrm H}$ which 
was tied to the same parameter for all sets and was measured to be $(2.434\pm0.008)\times10^{22}$\,cm$^{-2}$. 
We did not fit the conventional but more complicated blackbody plus power-law model because the 
addition of the extra blackbody component did not improve the goodness-of-fit significantly
for any of the \swift\ sets.  Although additional components are significant
for the \xmm\ and \cxo\ spectra, we opted to use a single component model because only \swift\ data
are used here to constrain the variability (see Section \ref{sec:variability}) and using a single-component
model simplifies the comparison of spectral properties.
The joint power-law fit to the sets of observations provided a Cash statistic of 32806 and a Pearson $\chi^2$ 
of 37191 for 33571 degrees of freedom. This corresponds to a reduced $\chi^2$ of 1.1.


Figure \ref{fig:flux} shows the 1--10\,keV absorbed flux and power-law index as a function of time resulting from the spectral fit to the sets. 
The mean and standard deviation of the flux for the PC mode sets are $4.0\times10^{-11}$\,erg\,cm$^{-2}$\,s$^{-1}$ and
$8.4\times10^{-13}$\,erg\,cm$^{-2}$\,s$^{-1}$, respectively. The PC mode photon index has a mean and standard deviation of
3.1 and 0.07. Thus for PC mode, the maximum variability allowed by $3\sigma$ confidence intervals 
(3 times the standard deviation divided by the average value)
is 6.3\% for the flux
and 7.5\% for the photon index. In WT mode, the mean flux is measured to be $4.1\times10^{-11}$\,erg\,cm$^{-2}$\,s$^{-1}$
with a standard deviation of $5.9\times10^{-13}$\,erg\,cm$^{-2}$\,s$^{-1}$ for a maximum variability of 4.3\%. For the
photon index, the mean and standard deviation are 3.2 and 0.04 for a maxiumum variability of 4.0\%.
Compared to the standard deviations measured for the individual observations above, the constraints
here from fitting the sets of observations improved as little as a factor of 1.2 (PC mode photon-index)
and as much as a factor of 3.5 (WT mode flux). A greater improvement is achieved with WT mode than PC mode,
which makes sense given that the individual WT mode observations have a large number of short ($\lapp2$\,ks) observations.

The probability of the data being constant can be estimated from the $\chi^2$ of each data set. 
For the PC mode observation sets, the reduced $\chi^2_\nu/\nu$ for the flux is 2.4/3 which corresponds to a 6.7\,\%
probability of being constant and for the power-law index the $\chi^2_\nu/\nu$ is 11/3 ($5.1\times10^{-5}$\,\% probability). 
The sets of WT mode observations have
a $\chi^2_\nu/\nu$ of 2.3/8 (2.1\,\% probability) for flux and a $\chi^2_\nu/\nu$ of 6.1/8 ($6.3\times10^{-6}$\,\% probability) 
for power-law index.
So, in both modes, the fluxes are consistent with being constant (i.e. within 3$\sigma$), although the power-law
indices are only consistent within a 5$\sigma$ tolerance. This could be due to
unknown systematic sources of error or possibly low-level spectral variations, possibly due to the neglected
blackbody component. Regardless, these variations are much lower than the $\sim30$\% previously claimed 
\citep{roz+05,cri+07,gri+07}. 

\subsection{Timing}

A phase-coherent timing solution for \src\ has been maintained using {\it Rossi X-ray Timing Explorer} since 1998; see \citet{dk13}.
In order to continue to maintain a timing solution for \src, we began a monitoring campaign using the \swift/XRT on 2011 July 28, overlapping with the {\it RXTE} campaign, until {\it RXTE}'s demise in December, 2011.
Monitoring observations were typically 2-ks long.
Barycentred events were used to derive a pulse time of arrival (TOA) for each observation. 
For a given observation, a TOA was obtained using a maximum likelihood (ML) method, as described by \citet{lrc+09} and \cite{snl+12}.
The ML method compares a continuous model of the pulse profile, derived from taking aligned profiles of all the pre-glitch \swift/XRT observations, and creating a template composed of the first five Fourier components.

These TOAs were fitted to a pulse arrival time model in which the phase, $\phi$, at time $t$ is given by:
\begin{equation}
\phi(t) = \phi_0+\nu_0(t-t_0)+\frac{1}{2}\dot{\nu_0}(t-t_0)^2.
\end{equation}
where $\phi_0$, $\nu_0$, and $\dot\nu_0$ are the phase, frequency, and frequency derivative of the pulsar
respectively at the reference epoch $t_0$.
This was accomplished using the TEMPO2 \citep{hem06} pulsar timing software package. 

\begin{figure}
\includegraphics[width=\textwidth]{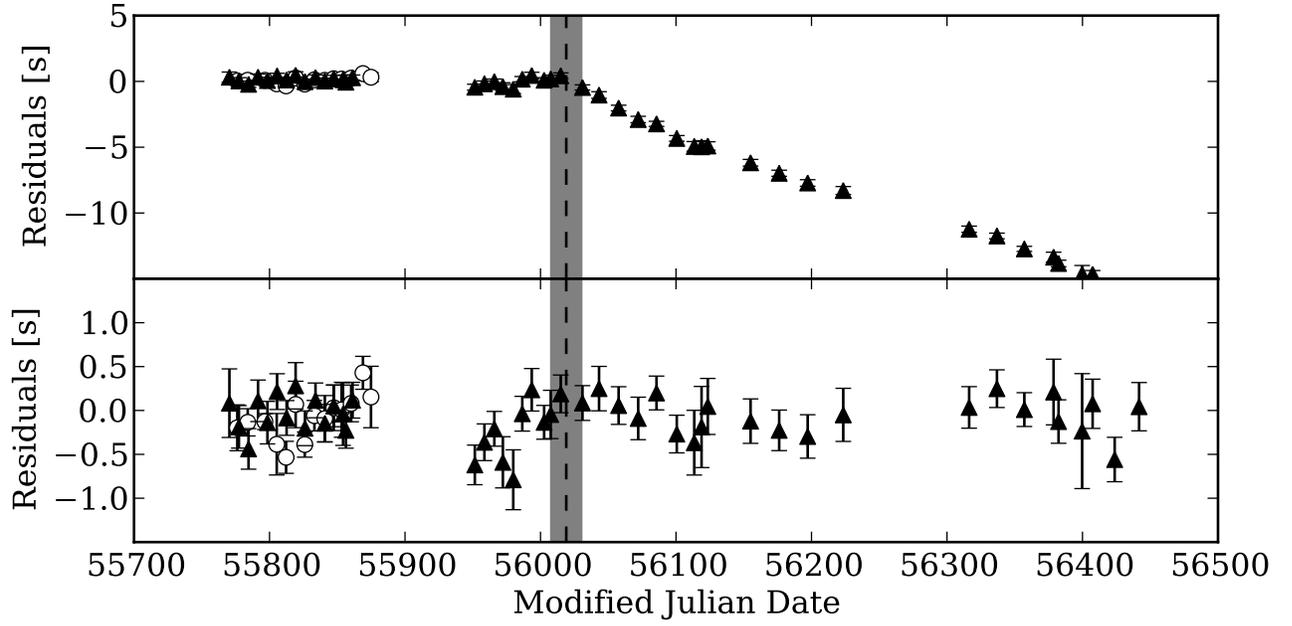}
\caption[Timing Residuals]{Timing residuals, the difference between the predicted and measured TOAs for 
the timing model shown in Table~\ref{ta:timing}. The top panel shows residuals before fitting for a glitch, 
and the bottom panel after. In both panels, open circles indicate data from {\it RXTE}, 
and black triangles indicate {\it Swift}. The vertical dashed line represents the glitch epoch and 
the grey band represents the uncertainty in that epoch.}
\label{fig:timing}
\end{figure}

In Figure~\ref{fig:timing} we show the timing residuals for \src\ starting on 2011 July 28 and show the overlap between the {\it RXTE} and {\it Swift} monitoring epochs.
The data are well fit by a single spin frequency and frequency derivative (Table \ref{ta:timing}).
However, we identified one notable timng event which we report as a new glitch.
The event occured within 11 days ($1\sigma$ uncertainty) of MJD 56019 with a decaying $\Delta\nu/\nu = (8.3\pm0.6)\times 10^{-7}$.
The glitch displayed an exponential recovery with a timescale of $111\pm15$ days, $2.6\pm0.3$ times longer than the $43\pm2$ day 
decay time of the other reported decaying glitch in the source \citep{dkg08}.
This glitch was accompanied by a $\Delta \dot{\nu}$ of $(1.4\pm0.3)\times 10^{-15}$\,Hz/s.

\section{Discussion}

In this paper, we have reported on the flux and spectral properties of \src\ over a $\sim10$ year period from 2003
to 2013. We show that there is no significant flux variability and that only low-level spectral variations are seen.
We have also presented an up-to-date timing solution and we report on
 a glitch that occured on MJD $\sim56019$. Below we compare our findings with
previous results and discuss the significance of the lack of variability in \src.

\subsection{Flux variability of \src}
\label{sec:variability}

In this work, we do not use measured flux and spectral properties between different
X-ray telescopes to constrain the variability of \src. 
This is because cross-calibration between instruments onboard
the \swift, \chandra, and \xmm\ telescopes is such that the flux and spectral index can differ by up to 20\% 
and 9\%, respectively \citep[e.g.][]{tgp+11}. 
As seen in Figure \ref{fig:flux}, the \xmm\ and \chandra\ observations are consistent with one another
within those tolerances.
Additionally, each \swift\ XRT mode (PC and WT) is considered separately, as the 
two modes are cross-calibrated only to within 10\% in flux (A. Beardmore, private communication).


Previous studies have claimed that \src\ displayed variability following glitches that occured between 2002 and 2005. 
Using a multi-component blackbody plus power-law model, 
\citet{gri+07} measure a low 1--10\,keV absorbed flux of $\sim3\times10^{-11}$\,erg\,cm$^{-2}$\,s$^{-1}$ for the 2003 \xmm\ observation and
2006 set of \swift\ observations. They measured a higher flux for the 2004 \chandra\ observation and the
2005 set of \swift\ observations. Their highest flux measured is $\sim4.5\times10^{-11}$\,erg\,cm$^{-2}$\,s$^{-1}$ for the 2005 \swift\ set.
This gives a total claimed variability of $\sim50\%$.

Because we used a single-component model in Section \ref{sec:spec_analysis}, the values in Figure \ref{fig:flux} are not
directly comparable to those in \citet{gri+07}. 
For the \swift\ data, additional spectral components do not significantly improve the fit. 
However, additional components for \chandra\ and \xmm\ observations do provide a much better fit and so here we apply 
a blackbody plus power-law model for direct comparison with \citet{gri+07}.
With a multi-component model we measure 1--10\,keV absorbed fluxes of
$(3.83\pm0.04)\times10^{-11}$\,erg\,cm$^{-2}$\,s$^{-1}$ for the \xmm\ spectrum and $(4.29\pm0.08)\times10^{-11}$\,erg\,cm$^{-2}$\,s$^{-1}$ 
for the \chandra\ observation.
This 11\% discrepency in flux between the two observations is within the 20\% cross-calibration error. 
As in \citet{gri+07}, we find that the temperature of the blackbody component of the model is consistent between the two observations and is $0.46\pm0.01$\,keV.
For the photon index we measure $2.63\pm0.03$ and $2.50\pm0.07$ for the \xmm\ and \chandra\ observations, respectively.
This compares to $\Gamma\sim2.8$ for both observations in \citet{gri+07}.
Reassuringly, the \chandra\ flux that we measure is higher than the \xmm\ flux and the \chandra\ spectral index is harder 
as found in cross-calibration studies \citep[e.g.][]{tgp+11}. 

In order to attempt to reproduce previous \swift\ results, for which only PC mode data were used \citep{gri+07}, we processed
the PC mode data from 2005-2007 without using exposure maps and without removing orbits with bad columns within
3 pixels of the center of the PSF. We found the same trend as in previous studies: the flux
of the 2005 set of observations was higher than the 2006 and 2007 sets and the flux of the 2007 set was slightly higher
than that of the 2006 set. We also observed an apparent correlation between the flux and power-law index. However, 
the level of variability in the three flux points was only about 30\% compared to $\sim50\%$ claimed in \citet{gri+07}.
Still, this is much higher than the $<10\%$ that we find in our more detailed analysis.

The lack of variability found here using soft X-ray imaging telescopes is consistent with what has been found 
by non-focusing telescopes in other regimes. Using \integral\ data, \citet{dkh08} found no significant variability in the hard X-ray flux
or spectral index for data spanning from 2003 to 2006. 
With \rxte, \citet{dkg08} found that the pulsed count rate showed evidence for only low-level variability ($<15\%$) and they concluded
that the glitches of \src\ appeared to be ``quiet", i.e. unassociated with significant changes in the radiative properites
of the magnetar.


\subsection{Radiative activity and Glitches in Magnetars}
\label{sec:glitches}

Radiative activity in magnetars is almost always associated with changes in timing behavior 
\citep[e.g. glitches or increased timing noise;][]{dk13}. 
Of the 26 known magnetars and magnetar candidiates, only five have long-term ($<$10\,yr)
phase-connected timing solutions that can be used to unambiguously detect glitches. These five magnetars are 
\meighteen, \mtwenty, \mohone, \mten, and \src. 
Of the three glitches each that have been detected from \mtwenty\ and \mten, five were radiatively loud, with 
the 2006 glitch of \mtwenty\ being the exception. The magnetars \meighteen, \mohone, and \src\ have not displayed any
significant flux increases associated with their glitches, although \mohone\ emitted short X-ray bursts near the epoch of its 2006
candidate glitch \citep[see][and references therein]{dk13}.

It is therefore clear that glitches are not always accompanied by radiative changes.
Because changes in the magnetosphere would likely manifest as pulse profile or flux variations, it seems more likely
that radiatively quiet magnetar glitches have their origin in the interior of the neutron star. 
If we assume that
radiatively quiet and loud glitches have the same origin, a mechanism must exist to allow magnetars to exhibit prompt
X-ray flux increases in some cases and no significant flux increases in others.

One possible way to achieve both radiatively loud and quiet glitches in an interior model
 is to vary the depth at which the glitch-inducing event occurs.
\citet{ec89a} showed that if energy is injected 
into the crust of a neutron star it can travel outward, and manifest as a prompt outburst, or travel inward 
and heat the core of the neutron star. The direction of travel depends on the size and depth of the energy deposition.
In the inward case, the heat is released slowly over a time scale of thousands of years. 
The flux decays of magnetars following prompt outbursts are indeed reasonably well modelled by crustal cooling 
\citep{let02,snl+12,akac13,skc13}. 
If the mechanism that causes glitches in magnetars injects energy at a shallow depth, a radiatively loud 
glitch would occur. 

An additional possible limit to the occurance of radiative outbursts from magnetars at glitch epochs
is the predominance of neutrino emission at high temperatures
in neutron star crusts \citep{ec89a,van91}. We expect neutron stars to have a limiting luminosity which
occurs when the emission of neutrinos dominates as a cooling mechanism over the emission of photons. 
We would thus expect the brightest magnetars to be unable to increase their luminosity beyond 
$\sim10^{35}$\,erg\,s$^{-1}$ \citep{td96a}. The five brightest magnetars, for which long-term
timing solutions are available, have luminosities $\sim10^{35}$\,erg\,s$^{-1}$ (though see below for
caveats on luminosity measurements). 
So, flux increases
for these magnetars should either not occur or be small. Indeed, of the five, only \mtwenty\ and \mten\ have displayed
significant flux increases at glitch epochs \citep{wkt+04,gk04,tgd+08} and those flux increases 
were much smaller than those from outbursts observed in fainter transient magnetars \citep[e.g.][]{icd+07,sk11}.

However, magnetar luminosities are not well constrained since the source distances are hard to determine.
There exist in the literature several disagreements in the distances to magnetars that lead
to a discrepency of up to a factor of $\sim30$ in luminosity (e.g.~see \citealp{akt+12} versus \citealp{dv06a} for \mten). 
Even when the distance is agreed upon there are discrepancies. For example, for \src, the only distance estimation
is from \cite{dv06a} but the 2--10\,keV luminosity has been reported to be as low as $4.2\times10^{34}$\,erg\,s$^{-1}$ 
\citep{ok13} and as high as $1.4\times10^{35}$\,erg\,s$^{-1}$ \citep{re11}.
From the model given by our best-fit mean flux and spectral indices, we get a 2--10\,keV unabsorbed flux 
of $3.9\times10^{-11}$\,erg\,cm$^{-2}$\,s$^{-1}$, which corresponds to a 2--10\,keV luminosity of $6.8\times10^{34}$\,erg\,s$^{-1}$,
closer to that listed in the magnetar catalog \citep{ok13}. Discrepancies such as this could be caused by
the difference in spectral models used or differences in the instruments used to measure the flux (as mentioned above,
X-ray detector cross-calibration can be discrepant up to 20\%).
We therefore cannot say conclusively whether magnetar luminosity is inversely correlated with the size of 
radiative activity as discussed here and as previously proposed in \citet{pr12}.

If we do assume that fainter magnetars are able to have larger flux increases
coincident with glitches, this suggests a rough luminosity order for the five brightest magnetars.
\src, \meighteen, and \mohone\ have experienced only radiatively quiet glitches whereas \mten\ and \mtwenty\
have shown significant flux increases during some (or all for \mten) of their glitches. That suggests that
\src, \meighteen, and \mohone\ are more luminous than the other two. 


Pulsars with higher B-fields are expected to be more luminous and have higher surface temperatures
than pulsars with lower magnetic fields because of energy deposition from the decay of their magnetic fields.
Indeed, it has been shown that high-B radio pulsars are systematically hotter than similarily aged pulsars 
with lower magnetic fields \citep{zkm+11,ozv+13}. 
We may also expect that magnetar-like activity in such sources 
could arise due to energy from the magnetic field being deposited at
shallow depths. Case in point, the high-B rotation-powered pulsar \kes\ displayed a magnetar-like outburst in 2006 \citep{ggg+08}.
In recent years, two magnetars, SGR~0418+5729 and Swift~J1822.3$-$1606, were discovered with magnetic fields lower 
than several high-B rotation-powered pulsars and have had clear X-ray outbursts 
\citep[though it is unknown whether or not they accompanied glitches;][]{ret+10,lsk+11}. 
It is thus becoming increasingly clear that high-$B$ rotation powered pulsars and magnetars
are related and form a spectrum of objects rather than two distinct groups. 
Therefore, the mechanism that causes X-ray outbursts at glitch epochs could
be active in all high-$B$ field pulsars.



\section{Conclusions}

We have presented an analysis of all of the \swift\ WT and PC mode data of \src\ in the period 2005--2013. We show that the 
maximum variability for both the 1--10\,keV X-ray flux and spectral index is constrained to $<10\%$. This is much
less than claimed by previous studies and is consistent with the flux being constant. We also report on a 
newly discovered glitch at MJD$\sim56019$ which has a fractional amplitude of
$\Delta\nu/\nu = (8.3\pm0.6)\times 10^{-7}$, typical of magnetar glitches.

The occurance of both radiatively quiet and loud glitches in magnetars, sometimes from the same source, shows that the mechanism
that causes these glitches must be able to produce prompt flux increases in some cases and no signifcant increases in others.
Here we have discussed the possibility that the glitches originate internally to the neutron star,
with the deciding factor the depth of the energy deposition associated with the
glitch. We note that these conclusions have been drawn from a sample of only five magnetars and therefore increasing
the number of magnetars for which we can unambiguously detect glitches would be beneficial in answering
the questions posed here.\\

We are grateful to the \swift\ team for their flexibility in the scheduling of the timing monitoring campaign of \src. 
We thank Marten van Kerkwijk, Andrew Cumming, Dave Tsang, and Kostas Gourgouliatos for helpful discussions.
V.M.K. holds the Lorne Trottier Chair in Astrophysics and Cosmology and a Canadian
Research Chair in Observational Astrophysics. R.F.A. recieves support from a Walter C. Sumner Memorial Fellowship.
This work is supported by NSERC via a Discovery Grant and an Accelerator Supplement, by FQRNT via the Centre
de Recherche en Astrophysique du Qu\'ebec, and by the Canadian Institute for Advanced Research.

\bibliographystyle{apj}
\bibliography{journals_apj,/homes/borgii/pscholz/Documents/papers/myrefs,modrefs,psrrefs,crossrefs}

\LongTables
\begin{deluxetable}{cccccccc}
\centering
\tablecaption{Summary of \swift\ observations of \src
\label{ta:obs}}
\tablewidth{0pt}
\tabletypesize{\tiny}

\tablehead{
\colhead{Sequence}&\colhead{Mode}&\colhead{Observation date}&\colhead{MJD}  &\colhead{Exposure time}&\colhead{Set}&\colhead{Set exp. time}&\colhead{Set counts}\\
\colhead{}        &\colhead{}          &\colhead{}          &\colhead{(TDB)}&\colhead{(ks)}         &\colhead{}   &\colhead{(ks)}         &\colhead{}
}

\startdata
00050701001 & PC & 2005-01-30 & 53400.2 & 2.3 & \multirow{5}{*}{2005}&\multirow{5}{*}{23.1}&\multirow{5}{*}{10947}\\
00050702001 & PC & 2005-02-02 & 53403.0 & 4.6 \\
00050702002 & PC & 2005-02-23 & 53424.0 & 2.0 \\
00050701002 & PC & 2005-02-24 & 53425.1 & 11.9 \\
00050700006 & PC & 2005-03-23 & 53452.2 & 2.3 \\
\hline
00035318001 & PC & 2006-09-20 & 53998.4 & 2.7 & \multirow{2}{*}{2006}&\multirow{2}{*}{11.8}&\multirow{2}{*}{5081}\\
00035318004 & PC & 2006-10-09 & 54017.3 & 9.2 \\
\hline
00035318005 & PC & 2007-02-25 & 54156.3 & 1.3 & \multirow{7}{*}{2007}&\multirow{7}{*}{12.4}&\multirow{7}{*}{5537}\\
00035318006 & PC & 2007-02-28 & 54159.0 & 1.8 \\
00035318007 & PC & 2007-03-05 & 54164.9 & 2.3 \\
00035318008 & PC & 2007-03-13 & 54172.7 & 1.2 \\
00035318010 & PC & 2007-03-18 & 54177.8 & 2.0 \\
00035318012 & PC & 2007-03-23 & 54182.6 & 2.0 \\
00035318011 & PC & 2007-03-26 & 54185.4 & 1.7 \\
\hline
00035318013 & PC & 2008-02-23 & 54519.2 & 15.9 & \multirow{2}{*}{2008-PC}&\multirow{2}{*}{20.9}&\multirow{2}{*}{10169}\\ 
00090025001 & PC & 2008-05-13 & 54599.0 & 5.0 \\
\hline
00090057001 & WT & 2008-04-02 & 54558.0 & 3.0 & \multirow{7}{*}{2008-1}&\multirow{7}{*}{27.1}&\multirow{7}{*}{26586}\\
00090057002 & WT & 2008-04-03 & 54559.6 & 2.1 \\
00090057003 & WT & 2008-04-04 & 54560.5 & 3.2 \\
00090057004 & WT & 2008-04-08 & 54564.1 & 1.2 \\
00090057005 & WT & 2008-04-11 & 54567.0 & 3.8 \\
00090057006 & WT & 2008-06-05 & 54622.2 & 6.4 \\
00090057007 & WT & 2008-06-06 & 54623.4 & 7.5 \\
\hline
00090057008 & WT & 2008-08-13 & 54691.3 & 7.4 & \multirow{4}{*}{2008-2}&\multirow{4}{*}{16.4}&\multirow{4}{*}{15260}\\
00090057009 & WT & 2008-08-14 & 54692.1 & 1.4 \\
00090057010 & WT & 2008-10-03 & 54742.0 & 7.0 \\
00090057011 & WT & 2008-10-10 & 54749.7 & 0.6 \\
\hline
00090057012 & WT & 2009-02-06 & 54868.3 & 2.2 & \multirow{5}{*}{2009-1}&\multirow{5}{*}{42.9}&\multirow{5}{*}{42616}\\
00090057013 & WT & 2009-02-08 & 54870.1 & 3.8 \\
00090057014 & WT & 2009-02-15 & 54877.1 & 12.7 \\
00090057015 & WT & 2009-03-20 & 54910.0 & 16.5 \\
00090213001 & WT & 2009-04-26 & 54947.1 & 7.7 \\
\hline
00090213002 & WT & 2009-06-28 & 55010.9 & 8.7 & \multirow{4}{*}{2009-2}&\multirow{4}{*}{22.3}&\multirow{4}{*}{21748}\\
00090213004 & WT & 2009-09-02 & 55076.2 & 8.7 \\
00090213005 & WT & 2009-10-11 & 55115.0 & 4.9 \\
\hline
00090213006 & WT & 2010-02-03 & 55230.0 & 8.6 & \multirow{3}{*}{2010}&\multirow{3}{*}{23.8}&\multirow{3}{*}{23038}\\
00090213007 & WT & 2010-02-04 & 55231.7 & 5.4 \\
00090213008 & WT & 2010-03-25 & 55280.0 & 9.8 \\
\hline
00035318014 & WT & 2011-07-28 & 55770.3 & 0.9 & \multirow{15}{*}{2011}&\multirow{15}{*}{27.8}&\multirow{15}{*}{27744}\\
00035318015 & WT & 2011-08-04 & 55777.4 & 1.0 \\
00035318016 & WT & 2011-08-11 & 55784.0 & 2.3 \\
00035318017 & WT & 2011-08-18 & 55791.2 & 1.9 \\
00035318018 & WT & 2011-08-25 & 55798.3 & 2.0 \\
00035318019 & WT & 2011-09-01 & 55805.4 & 2.1 \\
00035318020 & WT & 2011-09-08 & 55812.5 & 2.2 \\
00035318021 & WT & 2011-09-15 & 55819.2 & 2.2 \\
00035318022 & WT & 2011-09-22 & 55826.1 & 2.3 \\
00035318023 & WT & 2011-09-29 & 55833.9 & 2.4 \\
00035318024 & WT & 2011-10-06 & 55840.8 & 2.3 \\
00035318025 & WT & 2011-10-13 & 55847.0 & 1.6 \\
00035318026 & WT & 2011-10-20 & 55854.3 & 0.9 \\
00035318027 & WT & 2011-10-22 & 55856.3 & 1.8 \\
00035318028 & WT & 2011-10-27 & 55861.1 & 2.0 \\
\hline
00035318029 & WT & 2012-01-25 & 55951.5 & 2.0 & \multirow{14}{*}{2012-1}&\multirow{14}{*}{21.0}&\multirow{14}{*}{21056}\\ 
00035318030 & WT & 2012-02-01 & 55958.3 & 2.1 \\ 
00035318031 & WT & 2012-02-08 & 55965.8 & 2.2 \\ 
00035318032 & WT & 2012-02-15 & 55972.2 & 0.3 \\ 
00035318033 & WT & 2012-02-22 & 55979.8 & 0.4 \\ 
00035318034 & WT & 2012-02-29 & 55986.4 & 2.2 \\ 
00035318035 & WT & 2012-03-07 & 55993.3 & 2.2 \\ 
00035318036 & WT & 2012-03-16 & 56002.6 & 2.2 \\ 
00035318037 & WT & 2012-03-21 & 56007.5 & 1.0 \\ 
00035318038 & WT & 2012-03-28 & 56014.9 & 1.6 \\ 
00035318039 & WT & 2012-04-13 & 56030.2 & 1.6 \\ 
00035318040 & WT & 2012-04-26 & 56043.1 & 1.5 \\ 
00035318041 & WT & 2012-05-10 & 56057.3 & 1.3 \\ 
00035318042 & WT & 2012-05-25 & 56072.0 & 0.5 \\ 
\hline
00035318043 & WT & 2012-06-07 & 56085.6 & 2.0 & \multirow{9}{*}{2012-2}&\multirow{9}{*}{10.8}&\multirow{9}{*}{11423}\\ 
00035318044 & WT & 2012-06-22 & 56100.2 & 1.5 \\ 
00035318045 & WT & 2012-07-05 & 56113.4 & 0.6 \\ 
00035318047 & WT & 2012-07-15 & 56123.5 & 1.1 \\ 
00035318048 & WT & 2012-08-16 & 56155.0 & 1.7 \\ 
00035318049 & WT & 2012-09-06 & 56176.1 & 2.0 \\ 
00035318050 & WT & 2012-09-27 & 56197.1 & 1.7 \\ 
00035318052 & WT & 2012-10-23 & 56223.4 & 0.3 \\ 
\hline
00035318053 & WT & 2013-01-24 & 56316.3 & 1.0 & \multirow{5}{*}{2013}&\multirow{5}{*}{7.1}&\multirow{5}{*}{7380}\\ 
00035318054 & WT & 2013-02-13 & 56336.8 & 1.9 \\ 
00035318055 & WT & 2013-03-06 & 56357.1 & 2.0 \\ 
00035318056 & WT & 2013-03-27 & 56378.6 & 0.7 \\ 
00035318057 & WT & 2013-03-31 & 56382.2 & 1.5  
\enddata
\end{deluxetable}

\begin{deluxetable}{lc}
\centering
\tablecaption{Timing Parameters for \src.
\label{ta:timing}}
\tablewidth{0pt}
\tablehead{
\colhead{Parameter}&\colhead{Value}
}
\startdata
Observation Dates & 28 July 2011 - 29 May 2013 \\
Dates (MJD) & $55 770.396 - 56 441.770$\\
Epoch (MJD) & $56 000.000$\\
Number of TOAs & $61$\\
$\nu$  (s$^{-1}$)  & $0.090 851 264(3)$\\
$\dot{\nu}$ (s$^{-2}$) & $-1.638(3) \times 10^{-13}$  \\
\hline
\multicolumn{2}{c}{Glitch} \\
\hline
Glitch Epoch (MJD)  & $56019(11)$\\ 
$\Delta \nu_d$ (s$^{-1}$)   & $7.5(5)  \times 10^{-8}$\\
$\tau_d$ (days) & $111(15)$ \\ 
$\Delta \dot{\nu}$  (s$^{-2}$)   & $1.4(3)  \times 10^{-15}$\\
RMS residuals (ms)  & $229.07$\\
$\chi^{2} / \nu$  & $64.94/55$
\enddata \\
Numbers in parentheses are TEMPO2 reported 1 $\sigma$ uncertainties.
\end{deluxetable}

%


\end{document}